# Proton transfer at subkelvin temperatures

Lukas Tiefenthaler,[a†] Siegfried Kollotzek,[a†] Andrew M. Ellis,[b] Paul Scheier,[*a] and Olof Echt[*ac]

† Both authors contributed equally

**Abstract:**
We demonstrate a novel method to ionize molecules or molecular clusters by proton transfer at temperatures below 1 K. The method yields nascent ions and largely eliminates secondary reactions, even for notoriously 'delicate' molecules. Protonation is achieved inside liquid helium nanodroplets (HNDs) and begins with the formation of $(H_2)_mH^+$ ions as the proton donors. In a separate and subsequent step the HNDs are doped with a proton acceptor molecule, X. Proton transfer occurs between X and the cold proton donor ions inside a helium droplet, an approach that avoids the large excess energy that is released if HNDs are first doped and then ionized. Mass spectra, recorded after stripping excess helium and hydrogen in a collision cell, show that this method offers a new way to determine proton affinities of molecules and clusters by proton-transfer bracketing, to investigate astrochemically relevant ion-molecule reactions at sub-kelvin temperatures, and to prepare $XH^+$ ions that are suitable for messenger-tagging action spectroscopy.

a   S. Kollotzek, L. Tiefenthaler, Prof. P. Scheier
    Institut für Ionenphysik und Angewandte Physik
    Universität Innsbruck
    Technikerstraße 25, 6020 Innsbruck (Austria)
    E-mail: Paul.Scheier@uibk.ac.at
b   Prof. A. M. Ellis
    University of Leicester
    Leicester, LE1 7RH (UK)
c   Prof. O. Echt
    Department of Physics
    University of New Hampshire
    Durham, NH 03824 (USA)
    E-mail: olof.echt@unh.edu

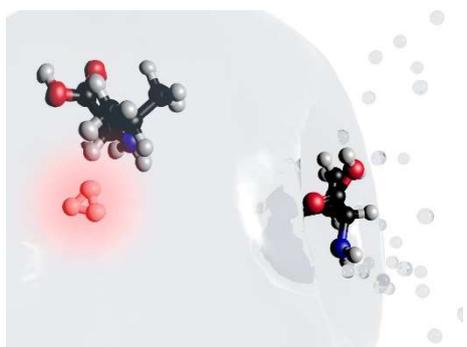

Table of Content Graphic
Doping cationic helium nanodroplets with $H_2$ produces proton donors. Subsequent doping with proton acceptors initiates gentle proton transfer at 0.37 K.

## Introduction

Proton transfer reactions (PTRs) are ubiquitous in chemistry. They are central to chemistry in liquid, hydrogen-bonded solvents and there are no better examples than the many PTRs that are essential in the chemistry of life.[1, 2] The importance of PTRs extends into the gas-phase, as demonstrated by PTR mass spectrometry, which is widely used to identify volatile organic compounds in the atmosphere with concentrations in the parts per trillion range.[3, 4] PTRs are also critically important in interstellar chemistry. About 15 % of the more than 200 molecules that have been identified by spectroscopic methods in the interstellar medium (ISM) are cations.[5] Most of these ions are protonated and presumably result from PTRs because their purported neutral precursors have also been identified in the ISM.[6]

PTRs are also prevalent when hydrogen-bonded clusters are ionized by electrons or photons.[7-10] If a homogeneous cluster composed of $p$ hydrogen-containing molecules X is ionized by electrons or photons, the most prominent cluster ions are usually protonated as a result of dissociative proton transfer from a cationic acid to the solvent moiety,

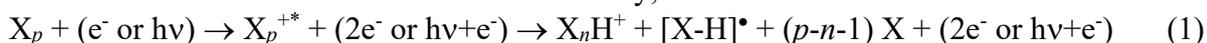
$$X_p + (e^- \text{ or } h\nu) \rightarrow X_p^{+*} + (2e^- \text{ or } h\nu + e^-) \rightarrow X_nH^+ + [X\text{-}H]^\bullet + (p\text{-}n\text{-}1) X + (2e^- \text{ or } h\nu + e^-) \quad (1)$$

where $0 < n < p\text{-}1$. The excess energy in the nascent $X_p^{+*}$ is usually much larger than the energy needed to eject the radical $[X\text{-}H]^\bullet$ and may also lead to the evaporation of several X molecules. Other competing intracluster dissociation channels often contribute significantly to the ion yield. For example, electron or photon ionization of bare $(CH_4)_p$ clusters not only produces $(CH_4)_n^+$ and $(CH_4)_nH^+$, but also $(CH_4)_nR^+$ with fragments R = $C_2H_2$, $C_2H_3$, $C_2H_4$, $C_2H_5$, $C_2H_6$, and $C_2H_7$ at the ~10 % level.[11-13]

Interest in interstellar chemistry has stimulated investigations of intracluster proton transfer and other ion molecule reactions at low temperatures and conditions that more closely resemble conditions in the ISM, where reactions are often catalyzed on the ice-covered surface of dust grains.[14] In one approach, a homogeneous or heterogeneous cluster is grown on small solid particles such as water-ice or argon-ice; their estimated temperatures are 100 K and 40 K, respectively.[15-17] In another approach, helium nanodroplets (HNDs) are formed in a supersonic expansion, passed through a pickup cell where they are doped with hydrogen-containing molecules, including $H_2O$, $NH_3$, formic acid, alcohols, ethers, or biomolecules, and subsequently ionized by electrons or photons.[18-25] The temperature of large HNDs (containing more than $\approx$100 helium atoms) is about 0.37 K.[26]

In spite of the low temperatures and the presence of the ice substrate or the helium matrix, proton transfer and other ion-molecule reactions are expected to be fast. Although the chemical reaction takes place in a nominally cold environment, it is initiated by the ionizing radiation, which enables excited-state proton transfer.[27] For ionization of bare clusters, the excitation energy equals, at the very least, the difference between the vertical and adiabatic ionization energy of the neutral precursor $X_p$. For ionization of a cluster embedded in a HND, the situation is even worse because the primary species formed by the ionizing radiation is $He^+$. Either this ion, or the stabilized intermediate $He_2^+$, will ionize the dopant by charge transfer. Either way, more than 10 eV will be released upon charge transfer to most embedded molecules.

Here we explore an alternative approach to ion formation in which the formation of the proton donor, $YH^+$, and the acceptor, X, are completely decoupled. The reaction

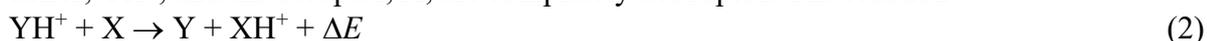
$$YH^+ + X \rightarrow Y + XH^+ + \Delta E \quad (2)$$

takes place between X and a fully thermalized donor $YH^+$ in a helium heat bath at 0.37 K. The reaction energy $\Delta E$ equals the difference between the proton affinities (PAs) of X and Y; it will be quickly dissipated into the helium matrix. In contrast to reaction (1), reaction (2) does not involve formation of a $[X\text{-}H]^\bullet$ radical.

Capture of further molecules X will result in growth of a protonated cluster $X_nH^+$ embedded in the HND. Each capture will release energy that is quickly dissipated and leads to shrinkage of the HND. In a final step, the doped HND passes through a collision cell filled with He at ambient temperature. Depending on the conditions chosen, collisions will result in the evaporation of all helium, yielding bare $X_nH^+$.

The small PA of $H_2$ (4.377 eV)[28] allows for proton transfer from $(H_2)_mH^+$ to a large variety of molecular clusters. In the present exploratory work we have studied proton transfer to $CH_4$ and valine, whose PAs are 5.633 and 9.438 eV, respectively.[28] The ionization processes of these molecules and their clusters have been well studied using a wide range of methods including electron, photon and chemical ionization, collisions with fission fragments and highly charged ions, and ionization of clusters embedded in helium droplets. The monomers show rich fragmentation patterns.[12, 29-33] For clusters, intracluster proton transfer dominates but several dissociative channels exist as well.[11, 13, 20, 29, 34-42] In contrast to mass spectra recorded with those conventional methods, the present approach leads to mass spectra in which the abundance of protonated ions $X_nH^+$ exceeds that of other product ions by orders of magnitude; they may very well be the sole products. Moreover, with suitably chosen pressure in the He collision cell, some $H_2$ molecules remain attached. The resulting ions are excellent candidates for action spectroscopy, where desorption of $H_2$ messenger molecules could be used to signal absorption of a photon.[43]

**Experimental Details**
A brief account of the experimental procedure is given here, with further details available in the Electronic Supplementary Information (ESI, S1). The experimental approach is illustrated in Fig. 1. Neutral HNDs are formed by supersonic expansion of pre-cooled helium into vacuum. The HNDs contain an average of about $2 \times 10^5$ or $9 \times 10^5$ helium atoms in experiments with methane or valine, respectively. The expanding beam is skimmed and ionized by electrons at 70 eV. The resulting ions, which may be multiply charged, are then accelerated into an electrostatic quadrupole bender. Ions that are transmitted have a specific size-to-charge ratio, $N/z$, with a spread of about 3 %.[44] In the present experiment the quadrupole bender was set to transmit ions with $N/z = 1.5 \times 10^5$ and $1.7 \times 10^5$ for methane and valine, respectively. The maximum charge state of droplets with a size-to-charge ratio $N/z = 1.5 \times 10^5$ equals $z_{max} = 17$.[44, 45] Multiply charged HNDs will, upon doping in the pickup cell and collisions with He in the collision cell, eventually end up as singly charged, doped HNDs. Details will be discussed in ESI S3.

Hydrogen gas is introduced into the vacuum chamber that houses the quadrupole bender. Successive capture of $H_2$ results in the growth of hydrogen cluster ions, $(H_2)_mH^+$, and evaporation of helium atoms from the droplet. In the pickup cell the doped droplets collide with proton acceptor molecules, which in the current experiments is either methane or D-valine.

The ions that exit the pickup cell pass through three regions in which they are guided by radiofrequency (RF) fields. The first one is filled with helium gas at ambient temperature and variable pressure, which will be referred to as the collision cell. Collisions with the background helium will cause evaporation of He atoms from the HNDs. If a sufficiently high number of collisions takes place collisional removal of $H_2$ molecules can also occur, leading to shrinkage of the cluster ions. The ions are then extracted into a commercial time-of-flight mass spectrometer equipped with a reflectron, see Fig. 1. For methane, the W configuration which involves two reflectrons was used.

**Results**
A mass spectrum of HNDs that were ionized by electrons, selected by their size-to-charge ratio $N/z = 1.5 \times 10^5$, and doped with first molecular hydrogen and then with methane is displayed in Fig. 2a. A spectrum recorded under the same conditions but without the addition of hydrogen is shown in Fig. 2b. These mass spectra are compared with previously published mass spectra, namely a mass spectrum recorded from neutral HNDs doped with $CH_4$ and then ionized by electron collisions (Fig. 2c)[35] and a mass spectrum of bare methane clusters seeded in He and electron ionized at 60 eV (Fig. 2d).[13]

The difference between the mass spectrum in Fig. 2a and the other mass spectra is dramatic. In Figs. 2b-d a multitude of peaks are seen. The spectrum in Fig. 2b shows peaks arising from $(CH_4)_nCH_5^+$ but also prominent additional peaks from $(CH_4)_nCH_x^+$, where $x = 1$, 3, or 4. The spectrum in Fig. 2c shows a large number of peaks that are not easily assigned while in Fig. 2d there are again many peaks and clear evidence of strong dissociative ionization processes. In contrast, the mass spectrum shown in Fig 2a is extremely simple, comprising of intense peaks, marked by triangles, from $(CH_4)_nCH_5^+$. (The satellite peaks at 2 u larger mass are caused by a water impurity, $(CH_4)_nH_3O^+$, rather than by $(CH_4)_nCH_5H_2^+$, see below). Consequently, we can conclude from this mass spectrum that the method we describe here almost entirely eliminates dissociative proton transfer reactions in the cases of methane and its clusters and delivers only the nascent proton transfer product.

Fig. 3 affords a similar comparison for valine (V) and its clusters. Clusters grown in charged HNDs ($N/z = 1.7 \times 10^5$) with and without $H_2$ pre-doping are displayed in Figs. 3a and 3b, respectively. Fig. 3c displays a mass spectrum recorded by electron ionization of neutral HNDs doped with valine reported by Denifl et al.[20] The series of weaker, equidistant mass peaks in that spectrum are due to $He_n^+$ ions that are ejected upon ionization.[44, 46] Data for the spectrum shown in Fig. 3d were read from a mass spectrum reported by Hu and Bernstein,[38] who photoionized a seeded supersonic expansion of valine in a helium/neon mixture. All spectra in Fig. 3 show prominent protonated valine clusters $V_nH^+$ at a nominal mass ($117n + 1$) u, which are marked by triangles. Asterisks mark fragment ion peaks at ($117n + 72$) u which correspond to loss of a carboxyl group COOH from $V_{n+1}^+$, the dominant fragmentation channel observed in previous studies of valine clusters.[10, 20, 37-42] The long progressions of mass peaks in Fig. 3a that follow the $VH^+$ and $V_2H^+$ ion peaks arise from these ions with multiple $H_2$ molecules attached. The pressure of helium in the collision cell was only 0.112 Pa, sufficient to strip the ions of all He atoms but not of all the $H_2$ molecules.[47]

The effect of increasing the collision pressure beyond the point where all He atoms are stripped from the doped HND is demonstrated in Fig. 4, for the case of $CH_4$ doped ions. In panel a $(CH_4)_nCH_5^+$ cluster ions at ($16n + 17$) u are marked by triangles and they are followed by a series of peaks due to attached $H_2$ molecules. In Fig. 4b the collision pressure was increased to 0.140 Pa, which greatly reduces the relative yield of the satellite peaks seen in Fig. 4a. Upon further increasing the collision pressure to 0.178 Pa the yield of peaks at ($16n + 19$) u relative to those at ($16n + 17$) u no longer decreases but the size distribution of $(CH_4)_nCH_5^+$ ions markedly shifts to smaller $n$, indicating collision-induced loss of $CH_4$. Why do the satellite peaks at ($16n + 19$) u not disappear? A closer look at the spectrum in Fig. 4a, reproduced in Fig. 4d with an expanded mass scale and logarithmic ordinate, provides the answer. Three clearly resolved mass peaks appear at nominally 83 u, namely a member of the ion series $(CH_4)_4CH_5(H_2)_x^+$ with $x = 1$, a member of $(CH_4)_3CH_5(H_2)_x^+$ with $x = 9$, and the ion $(CH_4)_4H_3O^+$ which is marked by a diamond. The two ion series quickly disappear when the collision pressure is increased but $(CH_4)_4H_3O^+$ stubbornly remains. As the PA of $H_2O$ is larger than that of $CH_4$ by 1.5 eV, evaporation of bare $CH_4$ is energetically preferred over evaporation of $H_2O$. Another ion peak in Fig. 4d, marked by an asterisk, is due to $(CH_4)_4CH_5^+$ containing one $^{13}C$ whose natural abundance is 1.07 %.

**Discussion**

We have presented a novel approach for the formation of cluster ions by proton transfer in superfluid helium droplets. In contrast to previous studies, proton transfer is not initiated by the addition of ionizing radiation nor by direct electron ionization, which would introduce large amounts of excess energy. Instead, the transfer takes place between a thermalized proton donor, in the present study $(H_2)_mH^+$, and a thermalized acceptor molecule X, resulting primarily in $XH^+$ with little or no ion fragmentation. Upon further capture of X molecules, a protonated cluster $X_nH^+$ will grow in the HND.

We will begin with a discussion of the various experimental steps before turning to the results presented in the previous section. Electron ionization of a pure HND initially results in $He^+$. The charge may move by resonant charge transfer,[48] but eventually the process is terminated by formation of a tightly bound $He_2^+$ ion. (If the HND is multiply charged it will contain $z$ $He_2^+$ that are well separated from each other). The proton donors are prepared by successive capture of $H_2$ molecules by the HND. For the first $H_2$, capture and charge exchange with one of the embedded $He_2^+$ ions

$$He_2^+ + H_2 \rightarrow He_2 + H_2^+ \tag{3}$$

will release approximately 6.88 eV of energy (details for this and the following reactions are to be found in the ESI S2), which will be quickly dissipated in the helium bath and lead to the evaporation of about 1.1×10⁴ He atoms (the binding energy per atom in bulk helium equals 0.616 meV[49]). Capture of another H$_2$ will lead to the reaction

$$H_2^+ + H_2 \rightarrow H_3^+ + H \qquad (4)$$

which will cause the evaporation of about 2900 He atoms. The large exothermicity of reaction (4) combined with the heliophobic character of the H atom[50] will, in all likelihood, cause its ejection (see the ESI S2 for details). (Alternatively, if the HND is multiply charged, capture of another H$_2$ may lead to a repeat of reaction (3) at another He$_2^+$. Ultimately, as a result of Coulomb explosion, all doped HNDs will be singly charged, as discussed in ESI S3).

Capture of further H$_2$ and growth of a (H$_2$)$_m$H$^+$ cluster ion will continue to release significant amounts of energy if $m$ is small.[51] The HND will have evaporated some 10% of its atoms by the time that the hydrogen cluster ions have grown to $m = 11$, where the fourth coordination shell closes.[52, 53] The energy release upon further H$_2$ capture will converge to 0.086 eV, the sum of the average collision energy, the internal energy of H$_2$, and the cohesive energy of bulk hydrogen.

Following growth of (H$_2$)$_m$H$^+$, the HNDs pass through a pickup cell filled with X = CH$_4$ gas or valine vapor. Capture of the first molecule leads to the reaction

$$(H_2)_mH^+ + X \rightarrow (H_2)_m + XH^+ \qquad (5)$$

which releases the difference between the PAs of CH$_4$ (5.633 eV) or valine (9.438 eV) and that of (H$_2$)$_m$, plus the sum of the average collision energy and internal energy of X. A maximum of about 2200 or 9600 He atoms will be evaporated upon capture and subsequent proton transfer for methane or valine, respectively.

Capture of further X will lead to growth of a protonated cluster ion

$$X_nH^+ + X \rightarrow X_{n+1}H^+ \qquad (6)$$

Dimer formation will cause the loss of about 680 and 3300 He atoms for methane and valine, respectively.[54,55] For large values of $n$, the corresponding values will be about 330 and 4100.

So far we have estimated the amount of energy released and the number of helium atoms lost upon collision of the charged HNDs with H$_2$ and either CH$_4$ or valine. The main purpose of that exercise was to show that, in spite of the large energies involved, the helium droplets have sufficient cooling power to remove the energy by evaporation. This is, of course, also evident from the mass spectra, which reveal (CH$_4$)$_n$CH$_5^+$ with $n$ up to 18 and V$_n$H$^+$ with $n$ up to 31. The protonated cluster can no longer grow once all the helium is evaporated. Furthermore, we find that collisions between the ions and He gas at ambient temperature are required to remove excess helium. H$_2$ molecules remain attached to the ions (see Figs. 3a and 4a) unless the He collision pressure is raised further.

The crucial point, however, is that the procedure leads to formation of protonated cluster ions as the dominant products, in stark contrast to conventional methods that also lead to abundant fragment ions. In our approach, capture of X and the proton transfer reaction (5) take place long after the formation of (H$_2$)$_m$H$^+$, thus guaranteeing vibrational relaxation of the proton donor prior to reaction. It takes roughly 0.5 ms for the HND to transit from the region of the quadrupole bender, where it is doped with H$_2$, to the pickup cell where it captures molecules X. This is ample time for thermalization of (H$_2$)$_m$H$^+$, and for the HND to return to its steady state temperature of 0.37 K.

Less obvious is the equilibration of X in reaction (5) or (6). The Landau velocity in superfluid HNDs is about 56 m/s.[56] Molecules whose incident velocity exceeds this critical velocity will quickly slow to 56 m/s. The radius of a HND containing 10⁶ atoms is about 20 nm, therefore the time between the capture of X and its encounter with the embedded proton donor will be a mere 0.4 ns if the incident particle moves on a straight trajectory to the embedded ion. However, molecular dynamics simulations reveal that the time is much larger than this naive estimate, because the captured particle will move on a rosetta-like trajectory.[57] For a HND radius of 20 nm, the average time will be about 10⁻⁷ s.[57, 58] Vibrational relaxation times in matrices span large time scales, but for most systems 100 ns will suffice for full vibrational relaxation.[59]

Reaction (6) involves another excited species, the nascent $X_{n+1}H^+$. Will this ion relax before encountering another X? The length of the pickup cell is 5 cm; the HNDs spend about 0.1 ms in the cell as they transit through. For an initial charge state of, say, $z \approx 10$ and an average number of 10 monomers per observed $X_nH^+$ ion, molecules X will be captured at a rate of roughly 1 μs$^{-1}$, providing ample time for vibrational relaxation between collisions.

Thus, reactions (5) and (6) involve reactant ions that have been cooled to 0.37 K. They offer the chance to grow protonated clusters $X_nH^+$ while largely avoiding intramolecular fragmentation of X, which is an intrinsic feature of other methods used for forming such ions (see reaction (1)). This difference is obvious when comparing Fig. 2a with the other panels in that Figure. For $(CH_4)_m$ clusters, conventional approaches not only lead to $(CH_4)_nCH_5^+$ but also several other ions that result from intracluster ion-molecule reactions, whose yields are comparable to that of $(CH_4)_nCH_5^+$.[11-13, 35, 36] In contrast, in Fig. 2a the only ions other than $(CH_4)_nCH_5^+$ result from a water impurity, as explained in the discussion of Fig. 4 above. Water impurities are, unfortunately, difficult to avoid in experiments involving large HNDs. Reducing the concentration of $H_2O$ in HNDs containing $10^6$ He atoms below 1 % would require the $H_2O$ presence in the expanding gas to be less than 0.01 ppm.

For valine (Fig. 3) the difference between conventional ionization methods and the current approach is not as obvious at first glance. Ionization of valine (V) molecules with the conventional methods produces a potpourri of fragment ions, most prominently [V-COOH]$^+$ due to cleavage of the C-C$_\alpha$ bond, but little or no V$^+$.[10, 30-33, 60] Electron ionization of valine clusters embedded in neutral HNDs[20, 41, 42] or photoionization of bare valine clusters[38] produces two major ion series, $V_nH^+$ (nominal mass $(117n + 1)$ u) and $[V_{n+1}$-COOH$]^+$ (nominal mass $(117n + 72)$ u), see Figs. 3c and 3d, respectively. Their yield obtained with conventional approaches exceeds 30 % that of $V_n^+$. Other methods, including desorption of valine films by fission fragments,[37] electrospray ionization,[39] and collisions of bare clusters with highly charged ions,[40] lead to similar results. This is not surprising, given that the energy required for formation of [V$_2$-COOH]$^+$ from $V_2^+$, predicted at the B3LYP/6-31+G$^{**}$ level of theory, is only 0.1 eV above that for formation of VH$^+$.[38]

The spectrum measured by our approach (Fig. 3a) shows groups of mass peaks near the positions of [V$_n$-COOH]$^+$ which are marked by asterisks. The insets in Fig. 3a offer expanded views. The relative yield of the most prominent mass peak in each group is about 0.5 % that of the corresponding $V_nH^+$. However, the mass of those peaks equals $(117n + 73)$ u rather than $(117n + 72)$ u for $n = 1$ and 2. Another prominent peak occurs at $(117n + 69)$ u. The nature of these peaks is unclear. Most likely they are due to impurities in the sample whose stated purity is 99.5 %; the specified concentration of foreign amino acids is below 0.3 %. Thermal decomposition products of valine in the pickup cell may also contribute. Amino acids have been found to decompose at temperatures above 450 K, slightly higher than the temperature (410 K) used in our experiment. The reported decomposition temperatures vary widely, from 434 to 588 K for valine.[61] Unfortunately the decomposition products have not been reported for valine but the loss of $CO_2$ would explain the peaks at $(117n + 73)$ u.[62] At any rate, the spectrum in Fig. 3a demonstrates that the yield of fragment ions due to loss of the carboxyl group COOH is much less than 1 %; they may very well be completely absent.

**Conclusion**

We have presented a novel scheme to form protonated molecules and molecular clusters. The method first involves formation of a proton donor, YH$^+$, in a helium nanodroplet. Subsequent addition of molecule X leads to proton transfer to form XH$^+$. If the partial pressure of X is raised sufficiently high, subsequent growth of a cluster $X_nH^+$ by capture of additional molecules is also possible. In this exploratory work the proton donor was $(H_2)_mH^+$ and the acceptor was either methane or valine. The temporal separation of the ionization events and their occurrence in a cold liquid environment ensures that dissociative proton transfer and other intramolecular dissociative reactions are largely avoided. Protonated cluster ions tagged with hydrogen molecules are a by-product if $(H_2)_mH^+$ cluster ions are chosen as the proton donor. These hydrogen-tagged cluster ions are well suited for recording action spectra of hydrogen-rich molecules of possible relevance to astrochemistry. Furthermore, the methodology described here may also prove useful to measure the proton affinity of molecular clusters by proton transfer bracketing using a variety of proton transfer reagents, YH$^+$. Finally, the low temperature of the helium matrix would have a marked effect on reactions that are exothermic but inhibited by small energy barriers, such as proton transfer from water to an amino group.[63]

**Acknowledgements**
This work was supported by the Austrian Science Fund (FWF Projects I4130 and P31149), and the European Regional Development Fund (EFRE K-Regio FAENOMENAL Project No. EFRE 2016-4).

**Conflict of interest**
The authors declare no conflict of interest.

**Keywords:** aggregation, helium, low-temperature chemistry, mass spectrometry, protonation

**Supporting Information**
S1   Experimental Details
S2   The Energetics of Ion-Molecule Reactions in HNDs
S3   The Fate of Multiply Charged HNDs
S4   Mass Spectra of Valine Clusters

**Author Contributions**
All authors contributed equally

**Notes**
The authors declare no competing financial interest.

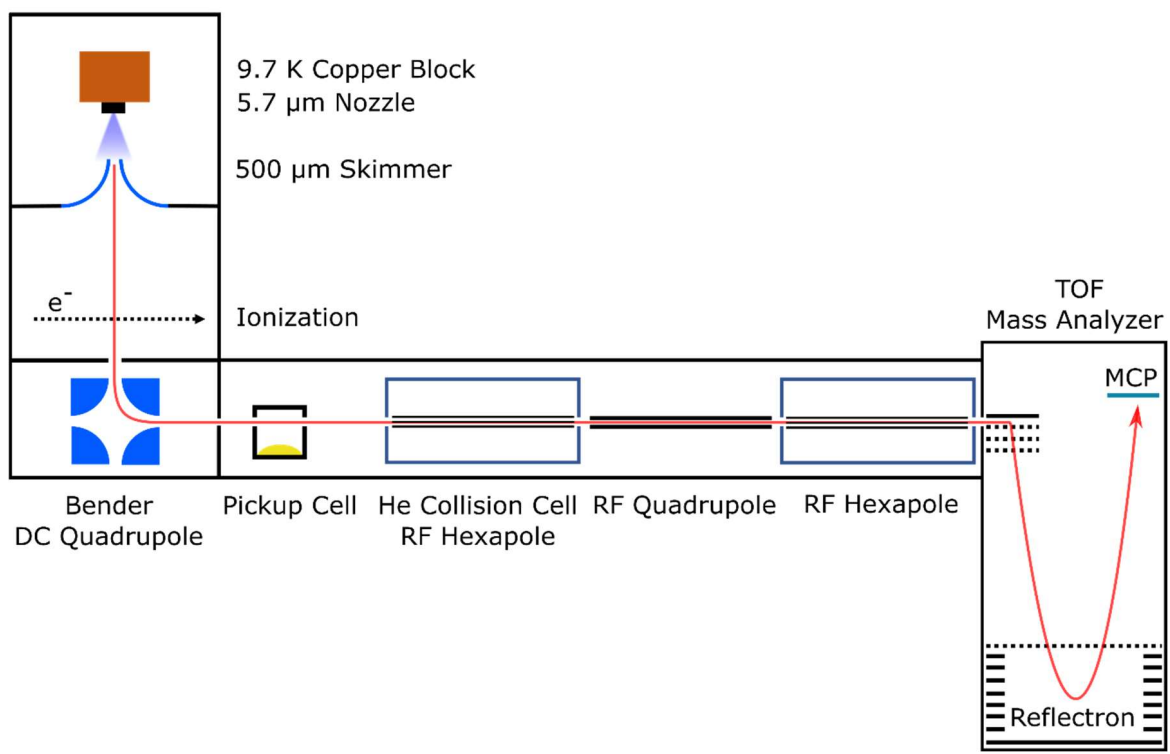

Fig. 1. The experimental setup used to initiate proton transfer between cold $(H_2)_m H^+$ and either methane or valine embedded in helium nanodroplets (HNDs). The ionized and size-to-charge selected HNDs are first doped with $H_2$ in the vacuum chamber housing the quadrupole bender, and then with proton acceptor molecules (either methane or valine) in the pickup cell. Helium and $H_2$ are then stripped from the embedded cluster ions by collisions with He gas at ambient temperature in the collision cell.

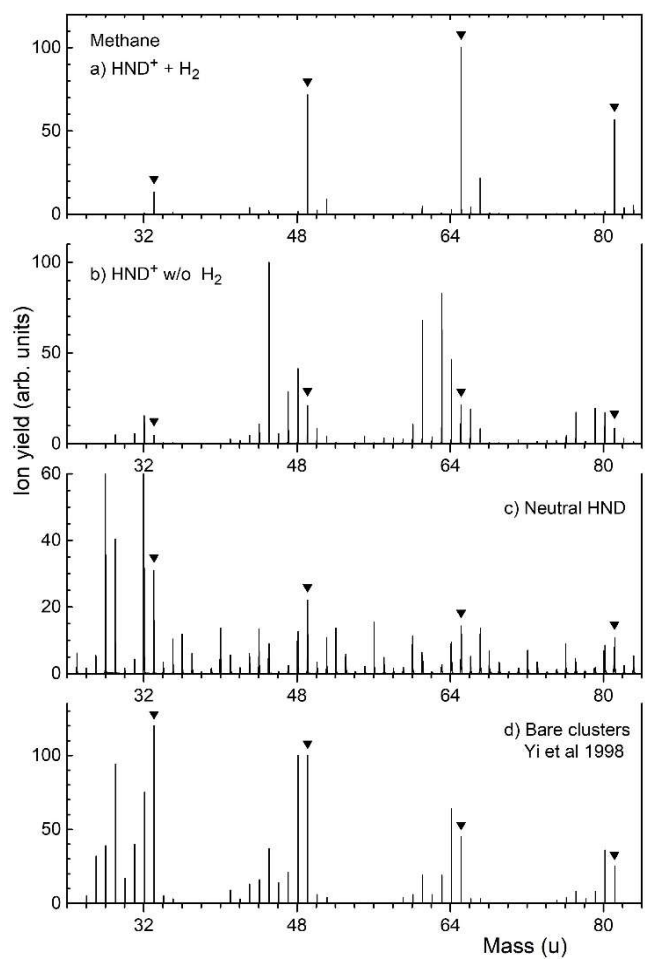

Fig. 2. Mass spectra of methane clusters. Panel a was recorded by doping charged HNDs first with $H_2$ and then with $CH_4$; panel b was recorded without pre-doping with $H_2$. The spectrum in panel c was recorded by electron ionization of neutral HNDs that were doped with $CH_4$.[35] The data plotted as a stick spectrum in panel d were read from a graph by Yi et al. who electron ionized bare methane clusters produced in a seeded expansion.[13] Triangles mark protonated methane clusters.

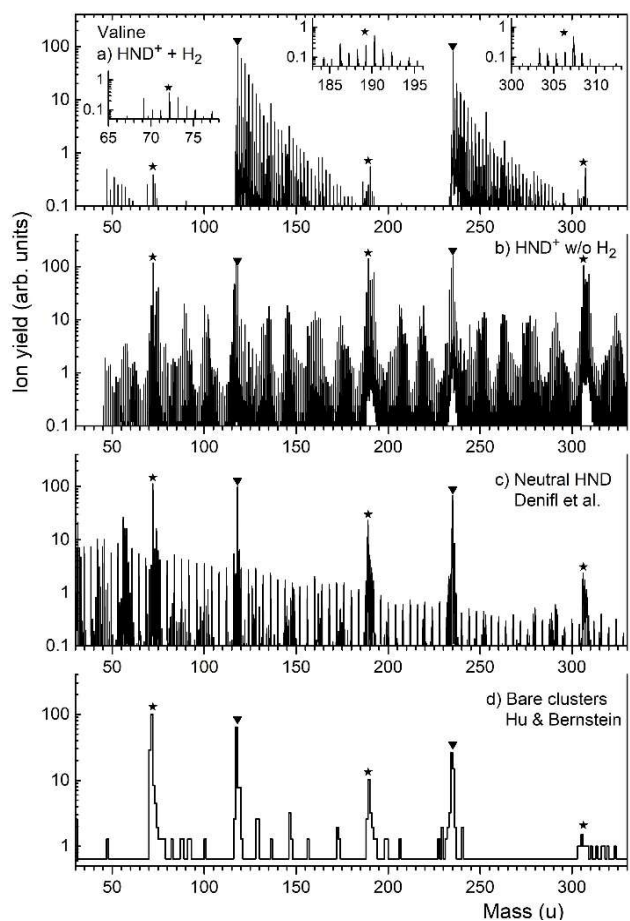

Fig. 3. Mass spectra of valine clusters. Panel a was recorded by doping charged HNDs first with $H_2$ and then with valine; panel b was recorded without pre-doping with $H_2$. The spectrum in panel c was recorded by electron ionization of neutral HNDs that were doped with valine.[20] The data plotted as a stick spectrum in panel d were read from a graph reported by Hu and Bernstein who photoionized bare valine clusters produced in a seeded expansion.[38] Triangles and asterisks mark the positions of $V_nH^+$ and $[V_n\text{-COOH}]^+$ ions, respectively.

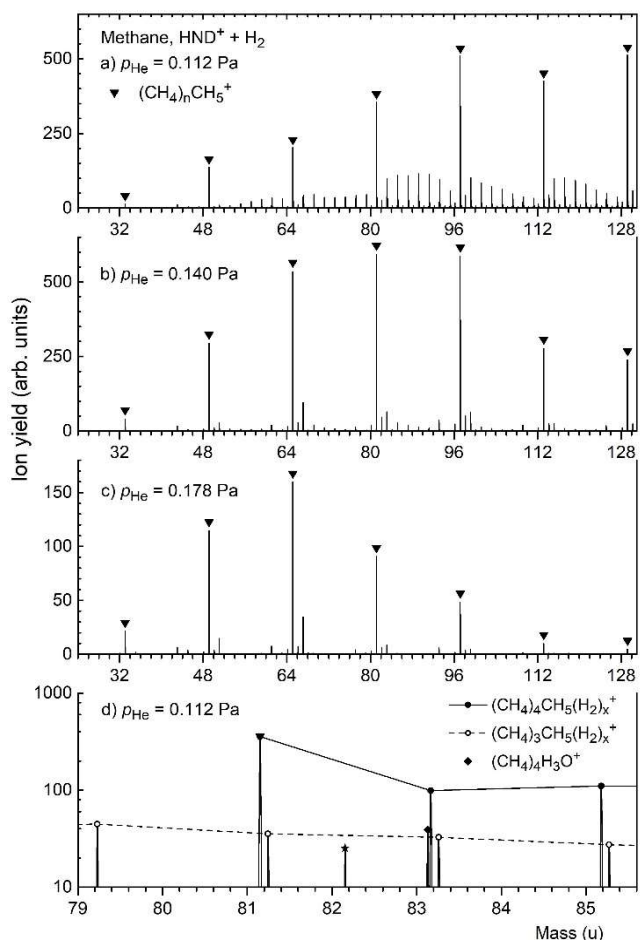

Fig. 4. Mass spectra of $(CH_4)_n$ clusters formed by doping charged HNDs first with $H_2$ and then with $CH_4$. The spectrum in panel a was recorded at relatively low collision pressure; protonated $CH_4$ clusters (marked by triangles) have not yet completely evaporated the attached $H_2$. Panels b and c: higher collision pressures result in the complete evaporation of $H_2$, but also of some $CH_4$. Panel d presents an expanded view of the spectrum recorded at low collision pressure; it reveals three different contributions to the group of mass peaks at nominally 83 u. The peak at 82 u marked by an asterisk is due to a $(CH_4)_4CH_5^+$ containing one $^{13}C$ isotope.

# Electronic Supplementary Information

# Proton transfer at subkelvin temperatures

Siegfried Kollotzek, Lukas Tiefenthaler, Andrew M. Ellis, Paul Scheier,* Olof Echt*

\* Electronic email: olof.echt@unh.edu, paul.scheier@uibk.ac.at

**Table of Contents**
S1  Experimental Details
S2  The Energetics of Ion-Molecule Reactions in HNDs
S3  The Fate of Multiply Charged HNDs
S4  Mass Spectra of Valine Clusters

## S1  Experimental Details

Neutral HNDs are formed by supersonic expansion of pre-cooled helium (Linde, purity 99.9999 %, stagnation pressure 20 bar) through a nozzle (5.7 μm diameter, temperatures 9.7 K for experiments with $CH_4$ or 9.2 K for experiments with valine) into vacuum. At these conditions the HNDs will contain an average of about $N = 2 \times 10^5$ helium atoms ($CH_4$) or $9 \times 10^5$ helium atoms (valine) but the size distribution will be broad.[1] The expanding beam is skimmed by a conical skimmer (Beam Dynamics, Inc.) and ionized by electrons at 70 eV. The electron emission current was 620 μA, resulting in multiple collisions between HNDs and electrons and highly charged droplets.[2] The resulting $He_N^{z+}$ ions are accelerated through a potential drop $\Delta V = 415$ V ($CH_4$) or 570 V (valine) into an electrostatic quadrupole bender. Ions that are transmitted have a specific size-to-charge ratio $N/z$. We chose $N/z = 1.5 \times 10^5$ and $1.7 \times 10^5$ for experiments with methane and valine, respectively.

Hydrogen gas is introduced into the vacuum chamber that houses the quadrupole bender (see Fig. 1 in the main text). The pressure, measured with a cold cathode ionization gauge (Pfeiffer model IKR 251), was $4.8 \times 10^{-4}$ Pa (all pressure values are corrected for the sensitivity of the ion gauge which is specified as 2.4 for $H_2$, 0.7 for $CH_4$, and 5.9 for helium). $H_2$ molecules colliding with a z-fold charged HND will be captured, resulting in the growth of z hydrogen cluster ions $(H_2)_mH^+$, and evaporation of helium atoms from the droplet.

Following the quadrupole bender the doped droplets pass through a pickup cell (length 5 cm) where they collide with proton acceptor molecules, either methane ($CH_4$, Linde, purity 5.5) or D-valine ($C_5H_{11}NO_2$, nominal mass 117 u, grade BioUltra, 99.5%, Sigma-Aldrich). The $CH_4$ pressure in the cell was $7 \times 10^{-4}$ Pa. Valine was vaporized in the cell at a nominal temperature of 410 K. At that temperature, its vapor pressure extrapolated from experimental values measured at higher temperatures is about $2 \times 10^{-4}$ Pa,[3] but the accuracy of our temperature measurement, and of the extrapolated values, is difficult to estimate.

The doped droplets that exit the pickup cell pass through three regions in which they are guided by a radio frequency (RF) field. The first one, a RF hexapole (length $L = 26$ cm), is filled with helium gas at 300 K. Collisions between the charged HND and the gas will cause evaporation of He. Further collisions will strip them of $H_2$ molecules and, eventually, lead to shrinkage of the dopant cluster ions. Multiply charged clusters will, at some point, undergo fission; their fate will be discussed in Section S3.

The following two ion guides, a quadrupole mass filter and a differentially pumped RF-hexapole collision cell, were not used in the present work. The ions were then extracted into a commercial time-of-flight mass spectrometer equipped with a double reflectron in W configuration and a microchannel plate (MCP) detector (Micromass Q-TOF Ultima mass spectrometer, Waters). $CH_4$ spectra were recorded in the W-mode and valine spectra in the V-mode. The mass resolution was 3000 at 264 u.

Mass spectra were evaluated by means of a custom-designed software that corrects for experimental artifacts such as background signal levels, non-gaussian peak shapes and mass drift over time[4] The routine takes into account the isotope pattern of all ions that might contribute to a specific mass peak by fitting a simulated spectrum with defined contributions from specific atoms to the measured spectrum in order to retrieve the abundance of ions with a specific stoichiometry.

For measurements with $CH_4$ (Figures 2a and 2b) the He pressure in the collision cell was $p_{He} = 0.178$ Pa, for valine (Figures 3a and 3b) it was $p_{He} = 0.224$ Pa.

## S2  The Energetics of Ion-Molecule Reactions in HNDs

In this section we will estimate the energies released in reactions 3 to 6, the secondary reactions initiated by this energy release, and competing reactions. The charged HNDs are mass-to-charge-selected by the quadrupole bender; their size-to-charge ratios are $N/z = 1.5 \times 10^5$ and $1.7 \times 10^5$ for droplets to be doped with $CH_4$ and valine, respectively. The main aim here is to show that these HNDs are sufficiently large to dissipate the energy released upon capture of multiple $H_2$ plus $CH_4$ or valine molecules. For the sake of clarity we will consider ions that are singly charged ($z = 1$); the fate of multiply charged HNDs will be addressed in Section S3.

The cohesive energy of bulk helium (isotopically pure $^4$He, to be exact) equals 0.616 meV/atom.[5] Thus we assume $D_{He} \sim 0.616$ meV for the evaporation energy of large HNDs, whether neutral or charged, doped or undoped. An energy of 1 eV released in a reaction will lead to the evaporation of about 1620 He atoms from the HND. The estimate neglects the average kinetic energy carried away by helium atoms that evaporate from a HND at 0.37 K, $E_{kin} = 3/2\ k_BT \sim 0.048$ meV. It also neglects the possible direct ejection of reaction products, or consequences of non-thermal loss of He.

A collision between a HND (neutral or charged) and a gas-phase molecule X in the pickup cell leads to capture of X with a near-100% probability.[6] X will quickly (within some 10 ns) move into the interior of the superfluid droplet ($H_2$, $CH_4$, or valine are not among the small number of species that are heliophobic). If the HND is charged or pre-doped, X will coagulate or react. The collision and subsequent coagulation/reaction will release energy $E^*$ that has three contributions:

$$E^* = E_{coll} + E_{int} + E_{reac} \tag{S1}$$

where $E_{coll}$ is the sum of kinetic energies of the collision partners in the center-of-mass reference frame, $E_{int}$ is the internal thermal energy (rotational + vibrational) energy of X, and $E_{reac}$ is the energy released upon coagulation or reaction.

The mass of the droplet which moves at drift speed $v_d$ through the pickup cell is much larger than the mass $m$ of X, i.e. the center of mass moves at speed $v_d$. $E_{coll}$ would equal $2\ k_BT$ if the droplet were at rest (where $k_B$ is the Boltzmann constant and $T$ the temperature of the gas). For a droplet moving through the thermal gas at speed $v_d$ one obtains[7]

$$E_{coll} = 2k_BT + m v_d^2/2 \tag{S2}$$

The speed $v_d$ depends on the temperature of the nozzle[1] and the electrostatic potential difference between the ionizer region and the pickup cell. The values are $v_d$ = 569 and 647 m/s for experiments with $CH_4$ and valine, respectively. Thus, in experiments with $CH_4$, each collision with $H_2$ or $CH_4$ (which are both at 300 K) adds $E_{coll}$ = 55 and 79 meV, respectively. In experiments with valine (temperature 410 K) the corresponding values are 56 and 323 meV.

The vibrational degrees of freedom in $H_2$ and $CH_4$ are essentially frozen at 300 K,[9] hence their internal energies $E_{int}$ equal approximately $k_BT$ = 26 and 1.5 $k_BT$ = 39 meV, respectively. For valine at 410 K, we estimate $E_{int} \sim 510$ meV from measurements involving crystalline valine.[10]

We proceed to estimate the energetics of specific ion-molecule reaction. Capture of the first $H_2$ molecule results in the reaction

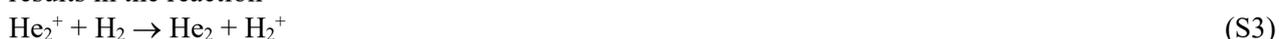

$$He_2^+ + H_2 \rightarrow He_2 + H_2^+ \tag{S3}$$

which will release a total of approximately 6.88 eV, the difference between the adiabatic ionization energies of $He_2$ (22.223 eV) and $H_2$ (15.426 eV)[11] plus $E_{coll} + E_{int} = 0.081$ eV. About $1.1 \times 10^4$ He atoms (equivalent to 7 % of the total) will be evaporated. Note that the energy release greatly exceeds the bond strength (2.7 eV) of $H_2^+$, hence ejection of an atomic H from the doped HND is conceivable. However, the bond length of $He_2^+$ is only 0.105 nm,[11] much less than that of $He_2$, hence vertical electron transfer from $H_2$ to $He_2^+$ results in a highly compressed $He_2$ molecule. Hence, ejection of a fast He atom is the more likely consequence of reaction S3, and much less than 7 % of He would be evaporated.

Capture of a second $H_2$ results in

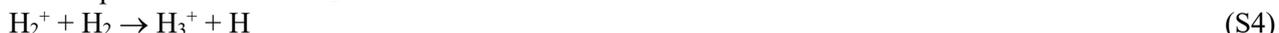

$$H_2^+ + H_2 \rightarrow H_3^+ + H \tag{S4}$$

which releases a total energy of about 1.8 eV, equivalent to the loss of approximately 2900 He atoms, or 2 % of the initial value. H is heliophobic; inside a HND it would reside in a large cavity,[12] and it will probably be expelled. The HND could conceivably quench the escape of H but experiments in which $H_2$-doped HNDs are ionized by electrons show a rather small signal of even-numbered $H_n^+$ cluster ions.[13] Tachikawa has studied vertical ionization of $H_2$ inside a small, bare $H_2$ cluster by ab-initio dynamics calculations.[14] He also finds that a fast H atom will be ejected.

Capture of further $H_2$ and growth of a $(H_2)_mH^+$ cluster ion will continue to release significant amounts of energies as long as $m$ is small. Nearly 2 eV (including the contributions from $E_{coll} + E_{int}$) will be released upon cluster growth from $m$ = 1 to 11.[15, 16] By that time the fourth solvation shell has been completed, and the energy release upon further captures will quickly converge to about 0.086 eV, the sum of the average collision energy, the internal energy of $H_2$, and the cohesive energy of bulk hydrogen (4.8 meV), causing the evaporation of a mere 140 He atoms per collision.

In the pickup cell, the $H_2$-doped HNDs will collide with X = methane or valine. The first collision

$(H_2)_mH^+ + X \rightarrow (H_2)_m + XH^+$                                                                       (S5)

releases the difference between the proton affinities (PAs) of $CH_4$ (5.633 eV) or valine (9.438 eV) and that of $(H_2)_m$, plus $E_{coll} + E_{int}$. $E_{int}$ can be estimated from the temperature dependences of the molar heat capacities of $CH_4$ and valine.[9,10] One finds $E_{coll} + E_{int} = 0.120$ and 0.830 V for $CH_4$ and valine, respectively. If $m = 1$, the total energy released in reaction S5 will be 4.50 and 5.21 eV for $CH_4$ and valine, respectively. This will cause the evaporation of about 2200 and 9600 He atoms for $CH_4$ and valine, respectively. For $m > 1$ the corresponding values will be smaller because the PA of $(H_2)_m$ exceeds that of $H_2$.

The energy released in the final reaction

$X_nH^+ + X \rightarrow X_{n+1}H^+$                                                                                                        (S6)

equals the sum of $E_{coll}$, $E_{int}$, and the energy released up cluster growth. The dissociation energy of $CH_4CH_5^+$ equals $E_{reac} = 0.30$ eV;[17] that of various amino acid proton-bound dimers is about 1.2 eV.[18,19] Hence $E^*$ equals about 0.42 and 2.03 eV for formation of the protonated methane and valine dimer, respectively; equivalent to the evaporation of some 680 or 3300 He atoms. For large values of $n$, $E_{reac}$ will gradually converge to the bulk cohesive energy which equals approximately 0.085 eV for $CH_4$ and 1.7 eV for valine.[3,17] The total energy release will then amount to about 0.18 eV for an added $CH_4$, and 2.3 eV for an added valine, equivalent to the loss of 330 or 4100 He atoms for $CH_4$ and valine, respectively.

In conclusion, large amounts of energy are released in some of the reactions considered, especially upon electron transfer from $H_2$ to $He_2^+$, and proton transfer from $(H_2)_mH^+$ to $X = CH_4$ or valine. Still, the number of helium atoms evaporated as a result of these reactions would be well below 10 % of the initial size of the HND. Furthermore, highly exothermic reactions are likely to lead to nonthermal ejection of reaction products, thus reducing the energy that needs to be dissipated by evaporation of He atoms.

Following the pickup cell the doped HNDs pass through the collision cell (length $L = 26$ cm) filled with He at 300 K. Each collision will transfer approximately $E^* = E_{coll} = 0.06$ eV, causing the evaporation of some 100 He atoms. The relations discussed in Section S3 may be used to estimate the path length $L_1$ after which a doped HND will have shed all of its He atoms. Over the remaining path $L_2 = L - L_1$ the ion will then shed excess $H_2$, producing bare $X_nH^+$ ions, and evaporate monomers X. This process is illustrated by the mas spectra shown in Fig. 4 of the main text.

**S3  The Fate of Multiply Charged HNDs**

Shrinkage of a $z$-fold charged HND will result in spontaneous charge separation (fission) once its size drops below its critical size for that charge state.[2] A droplet with an initial charge state $z_i$ close to the maximum number $z_{max}$ of charges that it can accommodate will, upon doping, quickly shrink to a size where it ejects a monocation complexed with a very small number of He atoms, plus a large, doped HND with $z_i-1$ charge centers. This process may occur several times in the pickup cell or, if $z_i \ll z_{max}$, not at all. But in the collision cell $N/z$ will definitely drop below the critical value of $5 \times 10^4$ for doubly charged droplets, and only singly charged doped HNDs will remain.

The primary product of electron ionization of a pure HND is $He^+$. In an undoped droplet the charge may move by resonant charge transfer but eventually the process is terminated by formation of a tightly bound $He_2^+$.[6] The high electron emission current in our experiment (620 μA) will result in multiple inelastic collisions between a given HND and the primary electrons, resulting in highly charged droplets. Charge states $z$ as large as 55 have been reported for droplets.[2] The $z$ $He_2^+$ ions will reside near the surface because of their mutual Coulomb repulsion. Upon multiple capture of $H_2$, the charge centers will transform into $(H_2)_mH^+$ where, on average, $m+1$ will equal the number of captured species divided by $z$. In the pickup cell, $n$ molecules $X = $ methane or valine will be added to each charge center where, on average, $n$ will equal the number of captured species divided by $z$.

The minimum size of a $z$-fold charged, undoped HND equals $N_z = 3.54 \times 10^4 z^{3/2}$.[2] This relation will also apply to doped droplets as long as the volume occupied by the dopant(s) is much less than the volume occupied by the $N$ He atoms. Once $N$ declines below $N_z$ (as a result of collisions with $H_2$ or with $X = CH_4$ or valine in the pickup cell), it will undergo spontaneous fission into a very small singly charged helium cluster containing fewer than about $10^2$ atoms, plus a $z-1$ fold HND with a size of approximately $N_z$.[2,20]

In the present experiment the quadrupole bender was set to transmit ions with $N/z = 1.5 \times 10^5$ and $1.7 \times 10^5$ for methane and valine, respectively. The corresponding maximum charge states are 17 and 23, respectively. For an illustration of the sequence of events, we consider a HND with $N/z = 1.5 \times 10^5$ in an initial charge state $z = 16$, containing $N = 2.4 \times 10^6$ He atoms. It becomes unstable with respect to fission once $N$ drops below $N_{16} = 2.27 \times 10^6$, i.e. after loss of $1.3 \times 10^5$ He atoms. This requires an energy release of 80 eV. Reaction S3 releases close to 7 eV. More than 80 eV will have been released even before all 16 $He_2^+$ charge centers have been converted to $H_2^+$, and the HND will undergo spontaneous fission into a 15-fold doped HND of size $N = $ $2.27 \times 10^6$. If the partial H$_2$ pressure is sufficiently high, the droplet will capture many more H$_2$, undergo reaction S3 a few more times, and reaction S4 many more times. The HND may undergo further fission events before it exits the vacuum chamber filled with H$_2$ gas.

The fate of a z-fold charged HND in the H$_2$ pickup cell may be estimated as follows: The capture cross section $\sigma$ of the droplet equals the hard-sphere value

$$\sigma = \pi(R_d + R_s)^2 = \pi\left(R_{He}N^{1/3} + R_s\right)^2 = \pi R_{He}^2\left(N^{1/3} + R_r^{-1}\right)^2 \tag{S7}$$

where $R_d$ is the radius of the droplet which contains $N$ He atoms, $R_{He}$ and $R_s$ are the effective radii of He and the scatterer H$_2$, and $R_r = R_{He}/R_s$ characterizes the size of He relative to that of H$_2$.

The droplet moves at speed $v_d$ through the scattering gas. $v_d$ may be computed from the speed of the undoped neutral HNDs which depends on the nozzle temperature, and the acceleration of the charged HNDs in the applied electrostatic field. In the present work, $v_d$ = 569 and 647 m/s for experiments with methane and valine, respectively.

The collision frequency $f$ equals

$$f = n_s v_r \sigma \tag{S8}$$

where $n_s$ is the number density of the H$_2$ scatterers which move at velocity $v$, and $v_r$ is the average relative collision speed in the center-of-mass system i.e. $|v-v_d|$ integrated over the surface of the droplet, all incident angles, and the thermal speed distribution of the scattering gas. The averaging can be simplified by noting that the asymptotic limit of $f$ equals $n\sigma v$ if the droplet moves slowly ($v_d \ll v$), and $n_s\sigma v_d$ if the droplet moves fast (because the droplet sweeps a volume $\sigma L$ within time $L/v_d$). We interpolate between the asymptotic limits with the expression

$$v_r \approx \left(v^2 + v_d^2\right)^{1/2} \tag{S9}$$

The number $dx$ of collisions with H$_2$ over a short path $dL$ may be written

$$dx = \frac{1}{\lambda}dL = \frac{f}{v_d}dL = n\sigma\frac{v_r}{v_d}dL \tag{S10}$$

where $\lambda$ is the mean free path of the HND moving through the scattering gas.

Each collision releases an energy $E^*$ (see eq. S1), resulting in the evaporation of $E^*/D_{He}$ helium atoms, hence the change in droplet size after $dx$ collisions equals

$$dN = -(E^*/D_{He})dx \tag{S11}$$

which can be combined with eqs. S7 and S10 to write

$$\left(N^{1/3} + R_r^{-1}\right)^{-2} dN = -n\pi R_{He}^2 \frac{v_r}{v_d}\frac{E^*}{D_{He}}dL \tag{S12}$$

Integration provides the relation between droplet size and path length provided $E^*$ is constant. If that assumption fails one has to compute $dL$ for each collision. For each initial charge state z, one can thus estimate the charge state and average size of the embedded (H$_2$)$_m$H$^+$ when the HND exits the vacuum chamber that houses the quadrupole bender. Note that each $z_1$-fold charged HND that enters this section produces one $z_2$-fold charged HND that exits. The small singly charged fragments that result from fission are of no interest because they do not have enough cooling power to capture additional molecules. Upon further collisions they will evaporate their constituents and end up as monomers or very small molecular ions and be lost on their way through the RF ion guides.

The fate of a $z_2$-fold charged HND containing $z_2$ (H$_2$)$_m$H$^+$ ions (where the average value of $m$ is known) in the pickup cell can be modeled as described above but now $v$, $n_s$ and $R_s$ now refer to the scatterer methane or valine.

In the collision cell the doped HND will lose all of its helium and along the way fission until $z$ = 1 (the minimum size of a doubly charged HND equals $N_2 = 1 \times 10^5$). It will also lose all or some of its H$_2$, and possibly some molecules X. The relations described above may be applied again, but now the quantities $v$, $n_s$ and $R_s$ refer to helium, and eq. S1 simplifies to $E^* = E_{coll}$. Furthermore, Eq. S7 needs to be modified when $N$ becomes small and the contribution of the embedded dopant(s) to the total volume of the HND is no longer negligible.

## S4 Mass Spectra of Valine Clusters

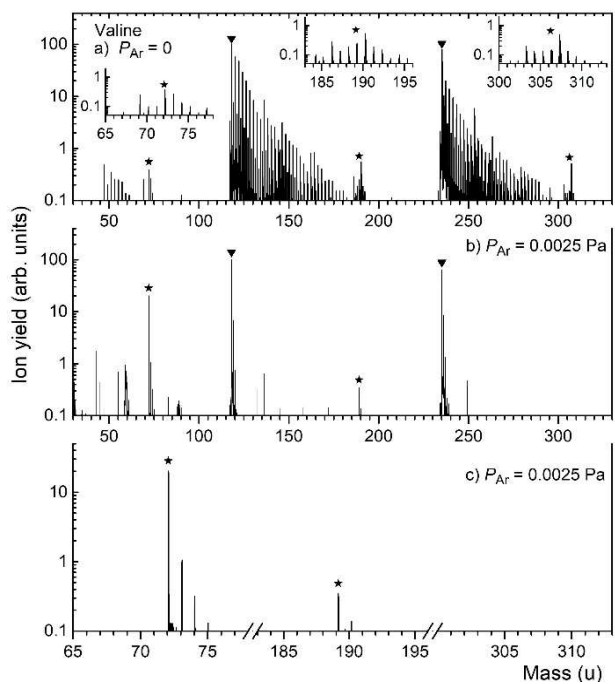

Fig. S1 Two mass spectra of valine clusters. The top panel reproduces the spectrum from Fig. 3a: Helium atoms were completely stripped from the doped HNDs, but some $H_2$ molecules remained attached to $V_nH^+$. Panel b displays a spectrum recorded after further dissociation was induced by collisions with argon gas in the last of the three ion guides (argon pressure 0.0025 Pa, collision energy 25 eV in the lab system). Triangles and asterisks mark the positions of $V_nH^+$ and $[V_n\text{-COOH}]^+$ ions, respectively. Panel c zooms into the same mass regions as the insets in panel a.